\documentclass[aps,prl,twocolumn,showpacs,superscriptaddress]{revtex4-1}

\usepackage{graphicx}
\usepackage{dcolumn}
\usepackage{bm}
\usepackage{amssymb}
\usepackage{colortbl}
\usepackage{dsfont}

\begin{document}

\title{Theory of carrier-mediated magnonic superlattices}

\author{J.P. Baltan\'as}
\email{baltanas@us.es}
\affiliation{Departamento de F\'{\i}sica Aplicada II, Universidad de Sevilla, E-41012 Sevilla, Spain}
\author{D. Frustaglia}
\email{frustaglia@us.es}
\affiliation{Departamento de F\'{\i}sica Aplicada II, Universidad de Sevilla, E-41012 Sevilla, Spain}

\date{\today}


\begin{abstract}
We present a minimal one-dimensional model of collective spin excitations in itinerant 
ferromagnetic superlattices within the regime of parabolic spin-carrier dispersion. 
We discuss the cases of weakly and strongly modulated magnetic profiles finding 
evidences of antiferromagnetic correlations for long-wave magnons (especially significant 
in layered systems), with an insight into the ground state properties. In addition, the
presence of local minima in the magnonic dispersion suggests the
possibility of (thermal) excitation of spin waves with a relatively well controlled wave length. 
Some of these features could be experimentally tested in DMS superlattices based on thin 
doped magnetic layers, acting as natural interfaces between (spin)electronic 
and magnonic degrees of freedom. 
\end{abstract}



\maketitle


\emph{Introduction.---} 
During the last decade, there has been an
increasing interest on the dynamics of collective spin excitations
(magnons or spinwaves) in magnetic materials at the nanoscale. Recent
efforts demonstrated that magnons can exhibit most of the
characteristic signatures of wave phenomena including the excitation
and propagation along wave guides, interference, reflection,
refraction, diffraction, focusing, and tunnelling, showing both
classical and quantum properties.  These findings established the
basis of the emerging field of \emph{magnonics}
\cite{NG-09,KDG-10,LHGM-11}, the aim of which is to exploit magnons to
carry and process information, among others.  The building blocks of
magnonics are magnetic superlattices for the design of bands that
determine the transport properties of magnons (and, specifically, band
gaps blocking magnon propagation).  Most developments have been
implemented in magnetic materials as ferrites and ferromagnetic alloys, 
where local magnetic moments interact
through direct exchange or dipolar coupling depending on the length
scale. Surprisingly, little attention has been paid in this respect to 
itinerant ferromagnets as diluted magnetic semiconductors (DMS)
\cite{JSMKM-06}. These are magnetic materials where local magnetic
moments are coupled through itinerant spin carriers in the absence of
direct coupling, allowing the electrical control of ferromagnetism
\cite{OCMOADOO-00}.  An example is GaMnAs, where magnetic
superlattices can be built by layering GaAs with modulated Mn-impurity
densities along the growth direction at the nanometer scale. These
structures were studied mainly in the context of spin-carrier
transport for applications in the field of \emph{spintronics} \cite{ZFD-04} as giant
magnetoresistance (GMR) by manipulating the (anti)ferromagnetic
coupling between Mn layers
\cite{JALM-99,GJG-08,CCLKBCLF-08,CLCYLKLF-10}. To our knowledge,
spin waves in magnetic semiconductor superlattices were only studied 
in the narrow-band limit under the action of direct impurity exchange, where 
carriers motion plays no role \cite{P-HPRRSK-11}.  
This family of hybrid materials present a natural advantage, acting as 
interfaces between (spin)electronic and magnonic degrees of freedom.
Certainly, this emerging aspects deserve more attention. 

Here we present a first step towards the theoretical study of
collective excitations in one-dimensional (1D) magnetic superlattices
mediated by itinerant carriers. It is based on a previous theory
originally developed for the modeling of ferromagnetism in uniformly
doped DMS \cite{KLM-00,FKM-04}, here extended for the study of
periodically modulated magnetic doping. The model accounts for dynamic correlations
between localized magnetic impurities and itinerant carriers in small
excitations from an ordered state, well beyond mean-field and RKKY
theories. We implement a path-integral formulation where carriers are
integrated out, obtaining an effective action for the impurity spins
expanded up to second order in the excitations amplitude.  We find a
number of spin-wave modes with distinguishing features as a
consequence of the periodic magnetic profile, showing signatures of
antiferromagnetic correlations absent in uniformly doped systems. In
combination with an external magnetic field, these could allow the
thermal excitation of spin waves of definite wavelength. We discuss
the cases of weakly and strongly modulated magnetic profiles in the
parabolic regime for carrier dispersion. This makes our study also
valid for 1D magnonic superlattices built upon the layering of 3D
systems (as, e.g., DMS superlattices). 

\emph{Model.---} We study a Kondo-like 1D model describing a system of
localized spins distributed periodically along the $z$-axis with spin
density ${\mathbf S}(z)$ coupled ferromagnetically to conduction-band
electrons. This is described by the Hamiltonian $H=H_{\rm kin}+H_{\rm
  Z}+H_{\rm ex}$, with contributions \cite{note-5}
\begin{eqnarray}
H_{\rm kin}&=&\int dz \sum_{\sigma=\pm 1}\hat{\psi}^{\dagger}_{\sigma}(z)
\Big(\frac{p^2}{2m}-\mu\Big)\hat{\psi}_{\sigma}(z), \\
\label{Hz}
H_{\rm Z}&=&\int dz \left [g \mu_{\rm B} {\mathbf B} \cdot {\mathbf s}(z)+g^* \mu_{\rm B}  {\mathbf B} \cdot {\mathbf S}(z) \right],\\
\label{Hex}
H_{\rm ex}&=&J \int dz~{\mathbf S}(z) \cdot {\mathbf s}(z).  
\end{eqnarray}
Here, $H_{\rm kin}$ accounts for the kinetic Hamiltonian of
conduction-band carriers [spinorial fields $\hat{\psi}_{\sigma}(z)$]
with effective mass $m$, introducing the chemical potential $\mu$ as a reference. 
An external magnetic field ${\mathbf B}$ contributes with a Zeeman energy $H_{\rm
  Z}$, Eq.~(\ref{Hz}), where we introduce the itinerant-carrier spin
density ${\mathbf s}(z)=\frac{1}{2} \sum_{\sigma \sigma'}
\hat{\Psi}_\sigma^\dag(z)\boldsymbol{\tau}_{\sigma
  \sigma'}\hat{\Psi}_{\sigma'}(z)$ with $\boldsymbol{\tau}$ the vector
of Pauli matrices. The coupling between itinerant carriers and local
spin is modeled by $H_{\rm ex}$, Eq.~(\ref{Hex}), with ferromagnetic
coupling constant $J < 0$ \cite{note-4}. This model simplifies considerably when the
itinerant-carrier density $n(z)$ is much smaller than the local-spin
density $N(z)$ \cite{STU91}, in which case we can treat $N(z)$ [and, 
consequently, ${\mathbf S}(z)$] as a continuous distribution 
where disorder is neglected by coarse
graining. This regime applies, among others, to bulk and
nanostructured DMS \cite{KLM-00,FKM-04}. 

By assuming positive g-factors $g$ and $g^*$, a field ${\mathbf B}=-B
\hat{z}$ tends to organize all spins parallel to the $z$-axis. This
serves as a mean-field reference state from which small spin
fluctuations are defined. By resorting to Holstein-Primakoff (HP)
bosonic fields $b(z)$ and $b^\dagger(z)$, the spin density ${\mathbf S}(z)$ can
be approximated in the small-fluctuation regime by $S^+ (z)\approx
b(z) \sqrt{2N(z)S}$, $S^-(z) \approx b^\dagger(z) \sqrt{2N(z)S}$, and
$S^z(z)=N(z)S-b^\dagger(z)b(z)$. This allows the introduction of a
coherent-state path integral representation for the partition function
in imaginary time $\xi$:
\begin{equation}
\label{Z}
\mathcal{Z}=\int\mathcal{D}[\bar{\Psi}\Psi]\mathcal{D}[\bar{\omega}\omega]\, 
e^{-\int_{0}^{\beta}d\xi\,\mathcal{L}[\bar{\Psi}\Psi,\bar{\omega}\omega]}
\end{equation}
with Lagrangian $\mathcal{L}=\int
dz\,[\bar{\omega}\partial_{\xi}\omega+\sum_{\sigma}\bar{\Psi}_{\sigma}\partial_{\xi}\Psi_{\sigma}
]+H[\bar{\Psi}\Psi,\bar{\omega}\omega]$. Here, we replace in $H$ the
fermionic-field (carrier) operators by Grassmann numbers
$\bar{\Psi}_\sigma, \Psi_\sigma$ and the HP bosonic (local-spin
fluctuation) operators by complex variables $\bar{\omega},\omega$
(where we omit the arguments $z,\xi$ for simplicity). By integrating
out the fermionic fields in (\ref{Z}) -- thanks to the bilinear
dependence of $\mathcal{L}[\bar{\Psi}\Psi]$--, we end up with an
effective picture in terms of local-spin degrees of freedom
$\mathcal{Z}=\int \mathcal{D}[\bar{\omega}\omega]
\exp(-\mathcal{S}_{\rm eff}[\bar{\omega}\omega])$ with an action
\begin{eqnarray}
\label{Seff}
\mathcal{S}_{\rm eff}&=&\int_{0}^{\beta} d\xi\,\int  dz\,[\bar{\omega}\partial_{\xi}\omega-g^*\mu_{\rm B}B(N(z)S-\bar{\omega}\omega)] \nonumber \\
&-&\ln\mathrm{det}(G^{-1}_{\rm MF}+\delta G^{-1}).
\end{eqnarray}
Here, the kernel $G^{-1}$ splits into mean-field ($G^{-1}_{\rm MF}$)
and fluctuating ($\delta G^{-1}$) contributions:
\begin{eqnarray}
\label{inverseG}
G^{-1}_{\rm MF}&=&\Big(\partial_{\xi}+\frac{p^2}{2m}-\mu\Big)\mathds{1}+\frac{\Delta(z)}{2}\tau_{z},\\
\delta G^{-1}&=&\frac{J}{2}\Big[-\bar{\omega}\omega\tau_{z}+\sqrt{2N(z)S}\Big(\omega\tau^{-}+\bar{\omega}\tau^{+}\Big)\Big],
\end{eqnarray}
where $\Delta(z)=J N(z)S-g\mu_{\rm B}B < 0$ is the (local) spin
splitting of the itinerant carriers. Notice that the dynamics of the
itinerant carriers is contained in the effective action, accounting
for the retarded interaction between local spins.

A non-interacting spin wave theory for the local spins is derived from Eq.~(\ref{Seff}) by
expanding $\mathcal{S}_{\rm eff}$ up to 2nd order in the bosonic
variables $\bar{\omega},\omega$. After dropping an irrelevant energy offset we
find
\begin{widetext}
\begin{eqnarray}
\label{Seff-swa}
\mathcal{S}_{\rm eff}[\bar{\omega},\omega]&=&
\int_{0}^{\beta}  d\xi\,\int dz\,\Big\{\bar{\omega}(z,\xi)\partial_{\xi}\omega(z,\xi)+g^*\mu_{B}B \ \bar{\omega}(z,\xi)\omega(z,\xi)-\frac{J}{2}[n^{\uparrow}_{\rm MF}(z)-n^{\downarrow}_{\rm MF}(z)] \ \bar{\omega}(z,\xi)\omega(z,\xi)\nonumber\\
&+&\frac{J^{2}S}{2}\int_{0}^{\beta}d\xi'\int dz'\,\sqrt{N(z)N(z')} \ G^{\uparrow}_{\rm MF}(z,\xi;z',\xi')G^{\downarrow}_{\rm MF}(z',\xi';z,\xi) \ \bar{\omega}(z',\xi')\omega(z,\xi)\Big\}.
\end{eqnarray}
\end{widetext}
Here we introduced the mean-field spin density $n^\sigma_{\rm MF}(z)$
and Green's function $G^\sigma_{\rm MF}(z,\xi;z',\xi')$ of the
itinerant carriers ($\sigma=\uparrow,\downarrow$) under the action of
an effective potential $U^\sigma(z)=\sigma \Delta(z)/2$ produced by
the combined action of the local-spin distribution $N(z)$ and the
magnetic field ${\mathbf B}$ \cite{note-3}. The 1st and 2nd terms of
Eq.~(\ref{Seff-swa}) are local in space, representing the mean-field
exchange field experienced by the local spins. The 3rd term is
non-local, instead, accounting for correlation effects due to the
response of itinerant carriers to local-spin reorientations.

The collective dynamics of the local spins is best understood by
studying the action $\mathcal{S}_{\rm eff}$ in Fourier representation
\begin{widetext}
\begin{equation}
\mathcal{S}_{\rm eff}[\bar{\omega},\omega]=\frac{1}{\beta}\sum_{j,n,m}\int \frac{dk}{2\pi}~\bar{\omega}(k+K_n,\nu_j)\left[D^{-1}(k,\nu_j)\right]_{nm}\omega(k+K_m,\nu_j),
\end{equation}
\end{widetext}
where the action's kernel is the inverse spin-wave propagator with
matrix elements $[D^{-1}]_{nm}$. These are ultimately determined by
the spectral decomposition of the periodic local-spin distribution
$N(z)=\sum_n N_n \exp(i K_n z)$, where $K_n= 2n\pi/z_0$ with $z_0$ the
magnetic superlattice constant. In the absence of disorder, a periodic
potential $U^\sigma(z)$ leads to the development of a carrier band
structure with 1st Brillouin zone defined by $-K_1/2 \le k \le K_1/2$,
with $k$ the carriers wave number, and lowest-band width $E_1 \approx
(\hbar^2/2m) (K_1/2)^2$ for almost-free carriers motion. Besides, the
expressions for $n^\sigma_{\rm MF}(z)$ and $G^\sigma_{\rm MF}(z,\xi;z',\xi')$ 
simplify conveniently by working within the
parabolic-band regime for a small majority-spin carrier Fermi energy
$E_{\rm F}=\mu+ |\Delta_0|/2 \ll E_1$ (with $\Delta_0=J N_0S-g\mu_{\rm
  B}B$ the carriers mean spin splitting), implying that the carriers
Fermi wave length is much larger than $z_0$. This eventually restricts our 
analysis to long-wave magnons.  

\emph{Elementary spin excitations.---} Close to uniformly doped DMS
\cite{KLM-00,FKM-04}, we find three different sets of
excitations. Collective modes with dispersion $\Omega(k)$ are
determined by studying the conditions under which $\det
D^{-1}(k,i\nu_j=\Omega)=0$. These modes organize in two branches at
different energy scales: soft modes $\Omega_{\rm soft} < x_{\rm s}
|\Delta_0|$ and hard modes $\Omega_{\rm stiff} \sim |\Delta_0|$, where
$x_{\rm s}=(\bar{n}^\uparrow_{\rm MF}-\bar{n}^\downarrow_{\rm
  MF})/2N_0S \ll 1$ is the ratio between carrier and local-spin mean
spin densities. Besides, a continuum of Stoner excitations
(corresponding to spin flipping in the carrier subsystem by
electron-hole transition) is determined by finding the $\Omega_{\rm
  S}(k)$ satisfying ${\rm Im} D^{-1} (k, i\nu_j)\neq 0$ after
analytical continuation $i\nu_j \rightarrow \Omega+i0$. In the
following, we focus our discussion on two study cases: weakly and
strongly modulated superlattices in the limit of vanishing magnetic
field ($B=0$). Moreover, we limit ourselves to a symmetric magnetic
profile $N(z)=N(-z)$, so that $N_n=N_{-n}=N_n^*$.

\emph{Weak modulation.---} We first consider the case of a finite
local-spin density $N_0$ perturbed by a weak harmonic modulation such
that $N(z)=N_0 [1+\alpha \cos(2\pi z/z_0)]$ with $\alpha \ll 1$. The
kernel reduces to $D^{-1}(k,i\nu_j)=-i \nu_j \mathds{1}- x_{\rm s}
\Delta_0 \mathds{1}+x_{\rm s} \Delta_0 I(k,i\nu_j) \mathds{M}$, where
each term derives from the corresponding 1st, 2nd and 3rd one of
Eq.~(\ref{Seff-swa}). Here, the integral factor
\begin{equation}
\label{I}
I(k,i\nu_j)= \frac{\Delta_0}{x_{\rm s} 2N_0S}\int \frac{dq}{2\pi} \frac{f(E_q^\uparrow)-f(E_{q+k}^\downarrow)}{i\nu_j+E_q^\uparrow-E_{q+k}^\downarrow}
\end{equation}
accounts for correlation effects due to the carriers response to local
spin reorientations, where $f(E_q^\sigma)$ is the Fermi distribution
for carriers of spin $\sigma$ and energy $E_q^\sigma=E_q+\sigma
\Delta_0/2-\mu$ with parabolic dispersion $E_q = (\hbar^2/2m)q^2$. We
evaluate Eq.~(\ref{I}) for zero temperature. Regarding $\mathds{M}$,
it is a symmetric diagonal-constant (Toeplitz) matrix with elements 1
and $\alpha/2$ along the 1st and 2nd diagonals, respectively.

\begin{figure}
\includegraphics[width=1.0\columnwidth]{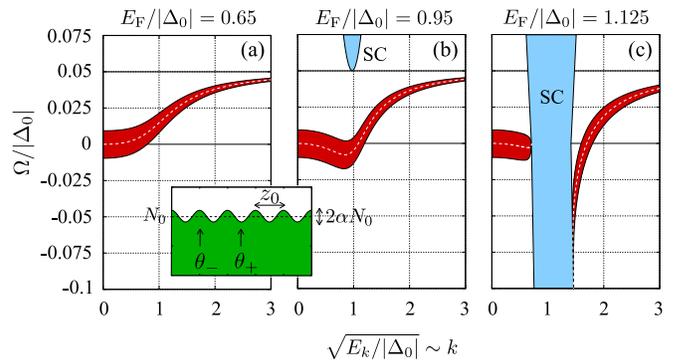}
\caption{
(Color online) Continuum of soft modes for
  fully [panels (a) and (b)] and partly [panel (c)] polarized carriers
  corresponding to \emph{weakly} modulated local-spin density as sketched in
  the inset ($\alpha=0.1$). In all cases, the lower and upper curves limiting the
  continuum are defined by the extreme values $\theta_{\mp}=0,\pi$,
  respectively. Note the development of minima as $E_{\rm
  F}/|\Delta_0|$ increases, occurring even in the case of uniform
  magnetic profile (dashed curve). The mean-field limit corresponds to 
  $\Omega/|\Delta_0| = x_{\rm s}=0.05$. In panels (b) and (c), the Stoner
  continuum (SC) lies between the curves $-\Delta_0+E_k\pm2\sqrt{E_{\rm
  F} E_k}$ for $E_{\rm F} \le |\Delta_0|$, and also between
  $-\Delta_0-E_k\pm2\sqrt{(E_{\rm F}+\Delta_0) E_k}$ for $E_{\rm F} >
  |\Delta_0|$.
} 
\label{fig-1}
\end{figure}

\begin{figure}
\includegraphics[width=1.0\columnwidth]{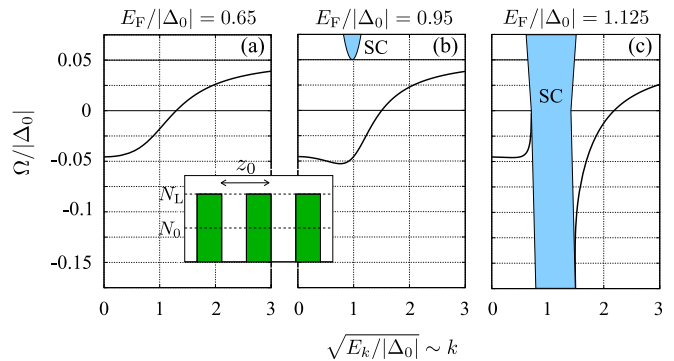}
\caption{
(Color online) Soft mode for
  fully [panels (a) and (b)] and partly [panel (c)] polarized carriers
  corresponding to \emph{strongly} modulated local-spin density as sketched in
  the inset ($\Gamma_0=0.5$). Note the negative excitation energy for small 
  $k$ and the development of a minimum as $E_{\rm
  F}/|\Delta_0|$ increases. The mean-field limit corresponds to 
  $\Omega/|\Delta_0| = x_{\rm s}=0.05$. The Stoner continuum (SC) in panels 
  (b) and (c) coincides with that of Fig.~\ref{fig-1}.
} 
\label{fig-2}
\end{figure}

We solve $\det [D^{-1}(k,\Omega)]=0$ by noticing that the Hermitian
$D^{-1}$ recalls a tight-binding Hamiltonian of an homogeneous,
infinite chain (in \emph{momentum} space) where Bloch's theorem applies: the
solution of the eigenvalue equation $D^{-1}(k,\Omega) | \theta \rangle
= \lambda(\theta) | \theta \rangle$ reads $\{ |\theta \rangle = \sum_n
\exp(i n\theta) |K_n\rangle,~\lambda(\theta)=-\Omega-x_{\rm s}
\Delta_0+x_{\rm s} \Delta_0(1+\alpha \cos \theta) I(k,\Omega)\}$,
where $0 \le \theta = 2\pi (z/z_0) \le 2 \pi$ for $0 \le z \le z_0$
and $|K_n \rangle$ is a spinwave state of wavenumber $K_n$. The
spinwave spectrum $\Omega(k)$ is then found by solving the equation
$\lambda(\theta)=0$. For each $\theta$ (indicating the location of the
excitation within each unit cell), we find two solutions corresponding
to soft and hard modes. A scanning over all values of $\theta$ yields
a continuum of excitations bounded by the curves defined by 
$\theta_\mp=0,\pi$. In Fig.~\ref{fig-1} we depict the
low-energy modes $\Omega_{\rm soft}$ together with the Stoner
continuum $\Omega_{\rm S}$ for fully ($E_{\rm F}\le |\Delta_0|$) and
partly ($E_{\rm F}> |\Delta_0|$) polarized carriers. For half
metallic carriers, we find a small-momentum (long-wave) dispersion
\begin{equation}
\label{soft-1}
\Omega_{\rm soft}=-\alpha \cos \theta \ x_{\rm s} |\Delta_0| +(1+\alpha \cos \theta) \ x_{\rm s} \left (1-\frac{4 E_{\rm F}}{3 |\Delta_0|} \right) E_k + {\cal O}(E_k^2).
\end{equation}
For large momenta (short wavelengths) we obtain the mean-field limit
$\Omega_{\rm soft} \rightarrow x_{\rm s} |\Delta_0|$. This is a consequence 
of the parabolic approximation for carriers dispersion: otherwise, periodic 
magnon dispersion is obtained. The results shown in Fig.~\ref{fig-1} (and 
Fig.~\ref{fig-2} as well) are then valid in the central region of the 
1st Brillouin zone. 

Several features stand out in Eq.~(\ref{soft-1}). The most important one is the
presence of negative-energy excitations ($\Omega < 0$). This means
that the reference state with all spins pointing along the $z$-axis is
actually \emph{not the ground state} \cite{note-1}. The latter must be
a complex state with lower magnetization, instead, developed by
long-range antiferromagnetic (AF) correlations present in the
system. Some AF signatures are already expected in \emph{uniform} 1D
systems \cite{MW-66}. The introduction of a weak modulation
represented by a finite $\alpha$ in Eq.~(\ref{soft-1}) provides an
additional source of AF correlations. This is illustrated by the
negative-energy excitations found in the neighborhood of $k=0$ for $0
\le \theta < \pi/2$ (doping \emph{hills}).  Besides, we find that
positive dispersions as a function of $E_k$ for a small $E_{\rm F} /
|\Delta_0|$ in Eq.~(\ref{soft-1}) turn into negative ones as $E_{\rm
  F} > (3/4)|\Delta_0|$ (reversing the sign of the spin-wave
stiffness and velocity), eventually leading to the development of a minimum with
negative energy (see Fig.~\ref{fig-1}). The latter is a purely 1D
characteristic: the position of the minima is independent of the
superlattice constant $z_0$ and only weakly dependent on the modulation
amplitude $\alpha$, persisting after setting $\alpha=0$ in
Eq.~(\ref{soft-1}) (uniform magnetic profile, depicted as the dashed
curve in Fig.~\ref{fig-1}, which also corresponds to
$\theta=\pi/2$). In addition, we find a set of finite (positive) energy
excitations around $k=0$ for $\pi/2 < \theta \le \pi$ indicating the
development of local magnetic anisotropies at the doping
\emph{valleys}.  We further notice that the spectrum can be lifted by
simply applying a magnetic field, turning negative-energy excitations
into positive ones by restoring fully-aligned spins as the ground
state. Interestingly, this opens a possibility to control thermal
excitation of magnons with relatively well defined wave number around
the minimum by a magnetic tuning of the gap.

\emph{Magnetic layering.---} We now consider the case of strongly
modulated local-spin density $N(z)= N_{\rm L} \Gamma(z)$ by
alternating magnetically doped and undoped layers, leading to a
periodic stepwise profile given by
\begin{equation}
\Gamma(z)=\Bigg\{
\begin{array}{rl} 
1 & \quad -\frac{z_{1}}{2}+n z_{0}<z<\frac{z_{1}}{2}+n z_{0} \\ 
0 & ~~~\text{otherwise},
\end{array}
\end{equation}
where $z_1$ and $N_{\rm L}$ are the width and local spin-density of
the magnetic layer, respectively, while $z_0$ is the superlattice
constant. We notice that a direct substitution of this profile on
Eq.~(\ref{Seff-swa}) could lead to unphysical situations as the
presence of local-spin excitations within undoped layers. To avoid this
problem, we redefine de HP-parametrization of the local-spin density
as $S^+ (z)\approx b(z) \sqrt{2N_{\rm L}S}~\Gamma(z)$, $S^-(z) \approx
b^\dagger(z) \sqrt{2N_{\rm L}S}\Gamma(z)$, and $S^z(z)=[N_{\rm
    L}S-b^\dagger(z)b(z)]\Gamma(z)$ \cite{note-2}. Let $\Gamma_n=(1/n
\pi) \sin(n \pi \Gamma_0)$ be the Fourier components of $\Gamma(z)$
with $0 <\Gamma_0=z_1/z_0 < 1$.  The kernel then reads
$D^{-1}(k,i\nu_j)=-i \nu_j \mathds{1}- x_{\rm s} \Delta_0
\mathds{M}_1+x_{\rm s} \Delta_0 I(k,i\nu_j) \mathds{M}_2$, with
$\Delta_0=JN_{\rm L}\Gamma_0 S$ (where $N_{\rm L}\Gamma_0=N_0$ is the
mean local-spin density) and $I(k,i\nu_j)$ defined in
Eq.~(\ref{I}). $\mathds{M}_1$ and $\mathds{M}_2$ are Toeplitz matrices
with elements $\Gamma_n$ and $\Gamma_n/\Gamma_0$ along the $n$th
diagonal, respectively.

To find the corresponding low-energy modes we proceed as in the
weakly-modulated case by studying the vanishing eigenvalues
$\lambda(\theta)=-\Omega + x_{\rm s} \Delta_0 (-1+
I(k,\Omega)/\Gamma_0)\sum_n \Gamma_n \cos(n \theta)=0$ of the
Hermitian $D^{-1}$. Here we find that the factor $\sum_n \Gamma_n
\cos(n \theta)$ in $\lambda(\theta)$ is nothing but $\Gamma(z_0
\theta/2\pi)=\Gamma(z)$ (by noticing that $\theta=2\pi (z/z_0)$ and
$\Gamma_n=\Gamma_{-n}$). This leads to three different kind of
solutions. The first one is determined by those $\theta$s satisfying
$\Gamma(z_0 \theta/2\pi)=1$. These correspond to a set of
\emph{degenerate inlaid modes} showing, in the half-metallic case, a
long-wave dispersion (see solid line in Fig.~\ref{fig-2} for the full
dispersion)
\begin{equation}
\label{soft-2}
\Omega_{\rm soft}=\left(1-\frac{1}{\Gamma_0}\right) \ x_{\rm s} |\Delta_0| + \frac{x_{\rm s}}{\Gamma_0} \left (1-\frac{4 E_{\rm F}}{3 |\Delta_0|} \right) E_k + {\cal O}(E_k^2).
\end{equation}
Here we find some features similar to those discussed in the
weakly-modulated case, including the presence of negative-energy
excitations and the switch to negative dispersion (with the eventual
development of a minimum) as $E_{\rm F}/|\Delta_0|$
increases. Interestingly, we also find a long-wave limit $\Omega_{\rm
  soft}(k=0)=(1-1/\Gamma_0)x_{\rm s} |\Delta_0|$, which
decreases as $\Gamma_0$ approaches zero. This means that AF
correlations become \emph{stronger} for thiner or widely separated
layers, taking the ground state away from that one with fully-aligned spin
configuration: reestablishing the reference state as the ground state
would require larger magnetic fields.  We secondly find a set of
spurious modes with vanishing energy for those $\theta$s satisfying
$\Gamma(z_0 \theta/2\pi)=0$, corresponding to undoped regions. These
solutions are of no physical relevance. Finally, we notice that the
series $\sum_n \Gamma_n \cos(n \theta)$ actually converges to 1/2
right at the interface between doped and undoped layers. This leads to
a new branch of modes, the dispersion of which is obtained by
replacing $x_{\rm s}$ by $x'_{\rm s} \equiv x_{\rm s}/2$ in
Eq.~(\ref{soft-2}). This means that spins placed at the interface feel
a local environment (here represented by $x_{\rm s}$) different from
those inlaid, lifting the long-wave excitation energy and lowering the
mean-field short-wave limit (eventually halving the band width with
respect to inlaid modes). These features are a consequence of the
particular mathematical properties of the stepwise
$\Gamma(z)$. However, more realistic profiles presenting gradual
interfaces shall develop a continuum of modes with similar
characteristics.

Additionally, we notice the development of Kohn-like anomalies
(divergence of spinwave group velocity $\sim \partial \Omega/\partial
k$) in the dispersion of low-energy modes for partly polarized
carriers, close to the points where meeting the Stoner continuum (see
Fig.~\ref{fig-2}).

{\em Conclusions.---} We present a minimal 1D model as a ``proof of concept" 
for the study of magnonic superlattices in itinerant systems, accounting for
back-action and correlation effects in the regime of parabolic
spin-carrier dispersion.  We discuss the cases of weakly and strongly
modulated magnetic profiles. The latter could be achieved in DMS
superlattices based on few nanometer thick doped magnetic layers
\cite{JALM-99,GJG-08,CCLKBCLF-08,CLCYLKLF-10}. This would allow to
reproduce some 1D characteristics of interest as negative magnon
dispersion and local minima in 3D systems, opening the door to thermal
excitation of spin waves with a relatively well defined wave number in a
controlled way.  The finding of negative excitation energies for
long-wave magnons indicate the presence of AF correlations, especially
strong for thin magnetic layers with $\Gamma_0 \ll 1$. This AF
signatures may arise from the effective coupling between distant
layers, and not necessarily between neighboring ones. The control of
spin-wave excitations in this context may require the application of
magnetic fields.  

{\em Acknowledgments.---} We thank A. Reynoso for useful comments. 
We acknowledge support from the Ram\'on y
Cajal program, from the Spanish Ministry of Science and Innovation's
projects Nos. FIS2008-05596, FIS2008-02873 and FIS2011-29400, and from
the Junta de Andaluc\'{\i}a's Excellence Project No. P07-FQM-3037.




\begin{thebibliography}{99}

\bibitem{NG-09}
S. Neusser and D. Grundler, Adv. Mater. {\bf 21}, 2927 (2009).

\bibitem{KDG-10}
V.V. Kruglyak, S.O. Demokritov and D. Grundler, J. Phys. D: Appl. Phys. {\bf 43}, 26400 (2010). 

\bibitem{LHGM-11} 
B. Lenk, H. Ulrichs, F. Garbs, and M. M\"unzenberg, Phys. Rep. {\bf 507}, 107 (2011). 

\bibitem{JSMKM-06} 
T. Jungwirth, J. Sinova, J. Ma\v{s}ek, J. Ku\v{c}era, and A.H. MacDonald,
Rev. Mod. Phys. {\bf 78}, 809 (2006).

\bibitem{OCMOADOO-00}
H. Ohno, D. Chiba, F. Matsukura, T. Omiya, E. Abe, T. Dietl, Y. Ohno, and K. Ohtani, Nature {\bf 408}, 944 (2000).

\bibitem{ZFD-04}
I. \v{Z}uti\'c, J. Fabian, and S. Das Sarma, Rev. Mod. Phys. {\bf 76}, 323 (2004). 

\bibitem{JALM-99}
T. Jungwirth, W. A. Atkinson, B. H. Lee, and A. H. MacDonald, Phys. Rev. B {\bf 59}, 9818 (1999).
 
\bibitem{GJG-08}
A. D. Giddings, T. Jungwirth, and B. L. Gallagher, Phys. Rev. B {\bf 78}, 165312 (2008).

\bibitem{CCLKBCLF-08}
J.-H. Chung, S. J. Chung, Sanghoon Lee, B. J. Kirby, J. A. Borchers, Y. J. Cho, X. Liu, and J. K. Furdyna, 
Phys. Rev. Lett. {\bf 101}, 237202 (2008).

\bibitem{CLCYLKLF-10}
Sunjae Chung, Sanghoon Lee, J.-H. Chung, Taehee Yoo, Hakjoon Lee, B. Kirby, X. Liu, and J. K. Furdyna,
Phys. Rev. B {\bf 82}, 054420 (2010).

\bibitem{P-HPRRSK-11}
M. Pavkov-Hrvojevi\'c, M. Panti\'c, S. Rado\v{s}evi\'c, M. Rutonjski, M. \v{S}krinjar, and D. Kapor, Solid State Commun. {\bf 151}, 1205 (2011).

\bibitem{KLM-00}
J. K\"onig, H. H. Lin, and A. H. MacDonald, Phys. Rev. Lett. {\bf 84}, 5628 (2000); J. K\"onig, T. Jungwirth, and A. H. MacDonald, Phys. Rev. B {\bf 64}, 184423 (2001).

\bibitem{FKM-04}
D. Frustaglia, J. K\"onig, and A.H. MacDonald, Phys. Rev. B {\bf 70}, 045205 (2004).

\bibitem{note-5}
To simplify the discussion, we neglect the interaction between free carriers. Some works have discussed its relevance on the magnon dispersion of uniformly doped low-dimensional magnetic semiconductors [see F. Perez, Phys. Rev. B {\bf 79}, 045306 (2009); C. Aku-Leh {\it et al.}, {\it ibid.} {\bf 83}, 035323 (2011);  F. Perez {\it et al.}, {\it ibid.} {\bf 83}, 075311 (2011); P. Shmakov {\it et al.}, {\it ibid.} {\bf 83}, 233204 (2011)]. The effects on 3D layered superlattices (as those proposed here) is yet unclear.

\bibitem{note-4}
For simplicity, we limit our study to the ferromagnetic case $J < 0$ expected to apply for n-doped systems. 
See Ref.~\onlinecite{FKM-04} for a discussion on the differences with the antiferromagnetic case $J > 0$ corresponding to p-doped systems. 

\bibitem{STU91}
M. Sigrist, H. Tsunetsugu, and K. Ueda, Phys. Rev. Lett. {\bf 67}, 2211 (1991).

\bibitem{note-3}
We disregard the particulars derived from considering local charge neutrality, since this affect both up and down spin carriers in the same way with little consequences on magnon dispersion at the lowest order. More relevant are, instead, the spin-dependent local potentials $U^\sigma(z)$ generated from the inhomogeneity of the doping. 

\bibitem{note-1}
The existence of gapless Goldstone modes (unidentified here) is not compromised 
since these are excitations referred to the ground state. 

\bibitem{MW-66}
N.D. Mermin and H. Wagner, Phys. Rev. Lett. {\bf 17}, 1133 (1966). 

\bibitem{note-2}
By following the construction procedure of coherent-state path integrals [see, e.g., J.W. Negele and H. Orland, \emph{Quantum Many-particle Systems}, Westview Press (1998), pp. 24 and 66-67], we notice that the term 
containing $\partial_\xi$ in Eq.~(\ref{Seff-swa}) is not affected by $\Gamma(z)$.    
 
\end{thebibliography}
\end{document}